\documentclass[a4paper,twoside,12pt]{JHEP3}
\usepackage{epsfig}
\usepackage{amssymb}
\usepackage{amsmath}
\def\roughly#1{\mathrel{\raise.3ex\hbox{$#1$\kern-.75em%
\lower1ex\hbox{$\sim$}}}}
\def\lsim{\roughly<}

\preprint{LYCEN-2004-15}

\title{$K \to \pi \nu {\bar \nu}$ from standard to new physics}

\author{Aldo Deandrea and Julien Welzel\\
Universit\'e de Lyon 1, Institut de Physique Nucl\'eaire,\\
4 rue E.~Fermi, F-69622 Villeurbanne Cedex, France\\  E-mail: 
\email{deandrea@ipnl.in2p3.fr, welzel@ipnl.in2p3.fr}}
\author{Micaela Oertel\\CEA DPTA-SPN, B.P. 12, 
F-91680 Bruy\`eres-le-Ch\^atel, France\\ E-mail: 
\email{micaela.oertel@cea.fr}}

\abstract{Flavour changing neutral current decays are a very sensitive
test of the standard model and its extensions. In particular the decay
$K \to \pi \nu {\bar \nu}$ constitutes a clean way to provide
constraints, independent of long distance effects. Motivated by the
recent experimental data of the E787 and E865 collaborations and by
the difference between the standard model prediction and data, we
consider in detail new physics scenarios such as the minimal
supersymmetric standard model and R-parity violating supersymmetry. We
begin with analyzing the impact of new measurements on the standard
model result obtaining $B(K^+ \to \pi^+ \nu {\bar \nu})=(8.18 \pm
1.22)\times 10^{-11}$.  Predictions for other rare kaon decays are
discussed, too. Our results allow to improve the limits on R-parity
violating couplings with respect to previous analyses.}


\begin{document}  

%
\section{Introduction}
%

The rare decay $K^+ \to \pi^+ \nu {\bar \nu}$ is a sensitive probe of quantum 
effects in the Standard Model (SM) and its extensions. We are mainly 
concerned in this paper with obtaining limits on the supersymmetric extensions 
of the Standard Model and in particular R-parity violating couplings. R-parity
violation allows tree level contributions to this decay and can therefore
be highly constrained. However the expected smallness of the R-parity 
violating couplings requires that the ``usual'' supersymmetric loop 
contributions are properly taken into account. 
Previous studies on upper bounds of R-parity violating couplings 
have been made without taking into account SM and supersymmetric 
1-loop contributions. Moreover previous limits do not take advantage of the
improved experimental results. Another reason to update the previous 
calculations both in the standard and the supersymmetric minimal model 
is the possibility to use the latest data in the flavour 
sector~\cite{S04,MTOP} and in related rare decays~\cite{KPINUNU,Ke3}. 
We also take the opportunity to correct a misprint in 
the neutralino contribution contained in the original 
publications. The impact of this change is quite small: the diagrams 
containing the neutralino contribute to the $\Delta F =1$ effective couplings 
together with small down-type Yukawa couplings in the squark mixing in the 
down sector. 

There are numerous classic papers on weak decays~\cite{Weak} and on
the analysis of the decay $K^+ \to \pi^+ \nu {\bar \nu}$ in the
context of the SM and supersymmetry with unbroken R-parity
\cite{classic}. We will follow the improved analyses of
\cite{BB99,DI,BRS,CI} and introduce in addition the breaking of
R-parity.

In the first part, we focus on the Standard Model neglecting all 
new-physics effects and we update in this context various related 
rare kaon branching rates.
In the second part, we take into account the one-loop supersymmetric 
contributions updating the same rare kaon branching rates.
Then in the third part we deduce more precise and realistic constraints 
on R-parity violating couplings involved in the $K^+ \to \pi^+ \nu 
{\bar \nu}$ decay. Unless otherwise stated all values are at 1-sigma level.

%
\section{The Standard Model}
%

Within the standard model the process $K^+ \to \pi^+\nu\bar{\nu}$ is
governed by the following effective Hamiltonian~\cite{BB99}
\begin{equation} 
H_{eff}= \frac{G_f}{\sqrt{2}} \frac{2 \alpha_e}{\pi \sin^2\theta_w} 
\sum_l \left( \lambda_c X^l_c +\lambda_t
X_t \right) \bar{s_L}\gamma^{\mu} d_L\;
\bar{\nu^l_L}\gamma_{\mu}\nu^l_L + h.c., 
\label{hamiltonian}
\end{equation}
where $\lambda_i$ are products of CKM~\cite{CKM} matrix elements: $\lambda_i =
V^*_{is} V_{id}$. $X_t$ contains the top contribution, and $X_c^l$ the
charm contribution for flavour $l$. QCD corrections have been calculated 
at the NLO level~\cite{BB99,QCD} and long-distance effects together with 
higher order electroweak effects are negligible~\cite{LongD}.
The top contribution does not depend on the lepton flavour since 
the lepton masses can be neglected with respect to the top mass. For the same 
reason the charm contribution for electrons and muons agrees and a function
\begin{equation}
P_c(X) = \frac{1}{3 \lambda^4} (2\, X_c^e + X_c^\tau)
\end{equation}
can be defined, see~\cite{BB99} for the explicit expression. 

The hadronic matrix element for the decay width can be related via
isospin to the experimentally well known decay $K^+ \to \pi^0 e^+ \nu_e$
~\cite{MP} and the branching ratio can be expressed as~\cite{BB99,DI}:
\begin{equation} 
BR^{SM}= \bar{\kappa}_+\left[ \left( \frac{Im(\lambda_t)X_t}{\lambda}
\right)^2 + \left( \lambda^4 P_c\,(X)\, \frac{Re(\lambda_c)}{\lambda} +
\frac{Re(\lambda_t)}{\lambda}X_t\right)^2\right] \; ,
\end{equation}
with 
\begin{equation}
\bar{\kappa}_+=r_+\; \frac{3 \alpha^2 (m_Z)\; BR(K^+ \to \pi^0 e^+ \nu_e)}
{2 \pi^2 \sin^4(\theta_w)} 
\end{equation}
$r_+ = 0.901$ is an isospin violation correction factor~\cite{MP} and
$\lambda = |V_{us}|$ in the Wolfenstein parameterization of the CKM
matrix~\cite{Wolfenstein}. The error in neglecting effects of higher
order in $\lambda$ in the improved Wolfenstein
parameterization is of the order
of $0.8\%$. Therefore we can safely neglect these corrections without
any significant loss in the predictions.

The branching ratio for the decay $K^+ \to \pi^0 e^+ \nu_e$ has
recently been measured with high 
statistics by the E865 collaboration~\cite{Ke3}. Their result, 
\[
BR(K^+ \to \pi^0 e^+ \nu_e)=(5.13\pm 0.15) \times 10^{-2}
\]
differs considerably from the most recent value of the Particle Data
Group~\cite{PDG04}, 
\[
BR(K^+ \to \pi^0 e^+ \nu_e)=(4.87\pm 0.06) \times 10^{-2} 
\]
which does not include yet the above mentioned result. 
We will use for our analysis
an average value for this branching ratio, where we
take into account the Particle Data Group
fit as well as the E865 result:
\[
BR(K^+ \to \pi^0 e^+ \nu_e)=(5.08\pm 0.13) \times 10^{-2} 
\]
This make the central value increase by 4.4\% and gives a larger error on all branching rates.
Then, $\bar{\kappa}_+=(7.97\pm0.20)\ 10^{-6}~.$

The CKM parameters are taken and updated from the recent fit of Stocchi~\cite{S04}
\begin{eqnarray*}
\mid V_{cb}\mid &=& 0.04135\pm0.0007 \\
\lambda&=&0.22385\pm0.00355\\
\bar{\rho}&=&0.190\pm0.046\\
\bar{\eta}&=&0.3485\pm0.0275
\end{eqnarray*}
These values have been obtained with the most recent value of the top
quark mass, $m_t({\bar m_t})=168.1\pm4.1\, \mathrm{GeV}$ in the
$\overline{MS}$ scheme, corresponding to the experimental value for
the pole mass of $m_t = 178.0 \pm 4.3\,\mathrm{GeV}$~\cite{MTOP}\footnote{
For the relation between the pole mass and 
top quark mass in the $\overline{MS}$ scheme see, e.g., Ref.~\cite{polems}.}

For the charm and top contribution we agree with Ref.~\cite{B04} and we use respectively :
\begin{eqnarray*}
P_c(X) &=& 0.39\pm0.07
\\
X_t &=& 1.529\pm0.042~.
\end{eqnarray*}

All these values can be combined to give our standard
model prediction for the branching ratio of $K^+ \to \pi^+ \nu {\bar
\nu}$ : 
\begin{equation}
BR^{SM}=(8.18 \pm 1.22)\times 10^{-11}.
\end{equation}
The central value differs slightly from the recent prediction by Buras
\textit{et al.}~\cite{B04} due to differences in the CKM parameters
resulting from different fits. The fit by Stocchi~\cite{S04} we used
includes some new E865 data and has been updated using the  $\overline{MS}$ 
top mass given above. The error on the branching ratio is
around 15\%, mostly due to the charm sector and to $\bar{\rho}$.
Buras \textit{et al.} \cite{B04} pointed out that the error in the
charm sector resulting from the renormalization dependence of the
function $P_c(X)$ could be reduced to 0.03 within a NNLO level
calculation of QCD corrections. Then the main uncertainty in the charm
sector arises from $m_c$.

The theoretical prediction is still compatible with the recent
experimental result for $K^+ \to \pi^+ \nu {\bar \nu}$, $BR = (1.47\
^{+1.3}_{-0.8})\ 10^{-10}$~\cite{KPINUNU}, even if the predicted central
value is half the observed value. It is too early to speculate about
the presence of new physics contributions as the experimental error is
still large, but it is interesting to note that possible new physics
effects should be of the same order as the SM ones in order to get the
measured central value. From these considerations one can conclude
that the rare decay $K^+ \to \pi^+ \nu {\bar \nu}$ is likely to play a
major role in the future both for constraining and discovering effects
beyond the standard model.

The effective Hamiltonian, Eq.~\ref{hamiltonian}, governs other
related rare kaon decays~\cite{B04}, too. We can thus easily obtain
predictions for the branching ratios of these decays:
\begin{eqnarray} 
BR^{SM}(K^0_L \to \pi^0 \nu \bar{\nu})&=& 
\bar{\kappa}_L\left( \frac{Im(\lambda_t)X_t}{\lambda}\right)^2 = 
(2.90 \pm 0.54)\times 10^{-11} \\
BR^{SM}(K^0_S \to \pi^0 \nu \bar{\nu})&=& 
\bar{\kappa}_S\left( \frac{BR^{SM}(K^+ \to \pi^+ \nu \bar{\nu})}
{\bar{\kappa}_+} -\frac{BR^{SM}(K^0_L \to \pi^0 \nu \bar{\nu})}
{\bar{\kappa}_L} \right) \nonumber \\
&=& (0.57\pm 0.09)\times 10^{-12} \\
BR^{SM}(K^0_L \to \pi^0 e^+ e^-) &=& (3.7 \pm 1.2)\times 10^{-11}
\end{eqnarray}
where
\begin{eqnarray}
\bar{\kappa}_L&=&\bar{\kappa}_+
\frac{\tau(K_L)}{\tau(K^+)}\frac{r_{K_L}}{r_{K_+}}=(3.49\pm0.09)\times 10^{-5}
\\
\bar{\kappa}_S&=&\bar{\kappa}_L\frac{\tau(K_S)}{\tau(K_L)}=(6.04\pm0.16)
\times 10^{-8}
\end{eqnarray}
and $r_{K_L}=0.944$ is the isospin correction for $K_L$~\cite{MP}.

Note that the process $K^0_L \to \pi^0 e^+ e^-$ is not determined
entirely by the short-distance contributions governed by the effective
Hamiltonian, but there are important long-distance effects. These have
been included in our estimate for the branching ratio following the
recent work in Ref.~\cite{IsidoriKL1,IsidoriKL2}. Using the average
value for $BR(Ke3^+)$, the result does not change significantly with
respect to that obtained in Ref.~\cite{IsidoriKL2}.  In contrast to
the decay $K^+ \to \pi^+\nu\bar{\nu}$, experimentally no process has
been observed so far corresponding to these decays. Hence on the
experimental side there exist only upper bounds on these branching
ratios which are orders of magnitude above the present theoretical
predictions~\cite{KL}. These decay processes are therefore not well
adapted to further constrain new physics scenarios. We only mention
these branching ratios in view of future data as theoretical
predictions can be easily obtained on the same footing as for
$K^+\to\pi^+\nu\bar{\nu}$.

%
\section{Supersymmetry}
%
In supersymmetric extensions of the standard model a large number of
particles carries flavour quantum numbers and can therefore contribute
to flavour changing neutral current processes~\cite{fcnc}. 
The $\Delta F=1$ decay
$K^+ \to \pi^+ \nu{\bar \nu}$ can be used in a model-independent way
to constrain new physics effects \cite{BRS}. There are no SUSY 
contributions at tree level, they start only at one-loop
order. At first sight the supersymmetric contributions thus seem
to be small, but as pointed out in Ref.~\cite{BRS} they can be of the
same order of magnitude as the standard model ones. At one-loop order
supersymmetric contributions to FCNC processes can be grouped into
three classes: the Higgs and W/quark exchanges, which contain as a
subclass the SM contributions; chargino and neutralino/squark
exchanges; and gluino/squark exchanges. The dominant supersymmetric
contributions are given by chargino/squark diagrams. In the following
part of this section we will detail these contributions following
Refs.~\cite{BRS,CI}. We will throughout this work use the nomenclature
``SUSY contributions'' for all supersymmetric contributions without
any R-parity violating ones as the latter will be discussed separately.

The determination of the supersymmetric contribution is based on the
same principle as in the SM case. There are two classes of diagrams:
penguins giving rise to an effective $Zsd$-vertex and box
diagrams. The effective Hamiltonian can be written in the same way as
in Eq.~\ref{hamiltonian} with $X_t$ replaced by $X_{new}=r_K
e^{-i\theta_K}X_t$~\cite{BRS}. $r_K$ and $\theta_K$ measure the amount
of new physics and are functions of masses of new particles and new
couplings. New physics effects proportional to $\lambda_c$ are
included in the new form of $X_t$.  The branching ratio for the decay
$K^+\to\pi^+\nu\bar{\nu}$ can then be expressed as
\begin{eqnarray} 
BR(K^+ \to \pi^+ \nu {\bar \nu})&=& \bar{\kappa}_+ \left( Im(C_l^{SUSY})^2 
+ Re(C_l^{SUSY})^2 \right) \nonumber \\
&=&\bar{\kappa}_+ \left\lbrace(r_K X_t \mid V_{cb}\mid^2 R_t)^2 
+(\lambda^4 P_c(1-\frac{\lambda^2}{2}))^2 \right.\nonumber \\ && \left. + 2 \mid
V_{cb}\mid^2(1-\frac{\lambda^2}{2})^2 \lambda^4 P_c X_t r_K R_t 
\cos(\beta + \theta_K) \right\rbrace 
\end{eqnarray}
where $\bar{\rho}=1-R_t \cos\beta$ and $\bar{\eta}=R_t \sin\beta$.

In writing the effective Hamiltonian in that way, we made some
assumptions. First of all we assumed that only one dimension six
operator contributes: $\bar{s_L}\gamma^{\mu}d_L\;
\bar{\nu^l_L}\gamma_{\mu}\nu^l_L$. At this stage we also assume
unbroken R-parity. The field content will be chosen minimal (MSSM
like). $X_{new}$ can then be written in the following form:
\begin{equation} 
X_{new}=X_t^{(SM)} + X_{H^{\pm}} + X_{\tilde{C}} + X_{\tilde{N}}~,
\end{equation}
showing the separate contributions in the loops of the standard model
particles, the charged higgses and up-quarks, charginos and
up-squarks (and charged sleptons in box diagrams), neutralinos and
down-squarks (and sneutrinos in box diagrams), respectively. The gluino
contribution (boxes and penguins) has been neglected. In the appendix
we recall all the formulae needed to compute the
SUSY contribution to the $d \to s\nu \bar{\nu}$ transition.  We have
corrected the misprints in $X_{\tilde{C}}$ and $X_{H^{\pm}}$ in Ref.~\cite{BRS} 
originally noticed by the authors of Ref.~\cite{CI}. Note also a misprint in the
neutralino penguins in the expressions in Ref.~\cite{BRS} which has
been corrected here (cf. appendix).

Due to the complicated flavour structure of supersymmetric theories, it
is often useful to diagonalize the mass matrices perturbatively 
via the mass-insertion approximation~\cite{MassInsertion}. 
In most cases it is sufficient to perform the single
mass-insertion approximation. Colangelo and Isidori \cite{CI} have
shown, however, that for the chargino/squark diagrams, it is necessary
to go beyond the single mass-insertion approximation in order to take
into account all possible large effects. Introducing the R-parameters
specifying the different mass insertions, 
\begin{eqnarray}
R^D_{s_L d_L}&=&\frac{(m_D^2)_{LL}}{\lambda_t m_{\tilde{d_L}}^2}=
\frac{(\delta^D_{12})_{LL}}{\lambda_t }
\\
R^U_{s_L d_L}&=&\frac{(m_U^2)_{LL}}{\lambda_t m_{\tilde{u_L}}^2}=
\frac{(\delta^U_{12})_{LL}}{\lambda_t }
\\
R^U_{s_L t_R}&=&\frac{(m_U^2)_{LR}}{\lambda_t m_t m_{\tilde{u_L}}} V_{td}=
\frac{(\delta^U_{23})_{LR}}{\lambda_t m_t}V_{td} m_{\tilde{u_L}}
\\
R^U_{t_R d_L}&=&\frac{(m_U^2)_{RL}}{\lambda_t m_t m_{\tilde{u_L}}} V_{ts}^*=
\frac{(\delta^U_{31})_{RL}}{\lambda_t m_t}V_{ts}^* m_{\tilde{u_L}}
\end{eqnarray}
the chargino/squark and the neutralino/squark contributions to
$X_\mathit{new}$ can be written as~
\cite{BRS,CI}:
\begin{eqnarray} 
X_{\tilde{C}}&=& C^0 + C_{LL} R^U_{s_L d_L} 
+ C_{LR} R^U_{s_L t_R}+ C_{RL} R^U_{t_R d_L}
\\
X_{\tilde{N}}&=& \tilde{N} R^D_{s_L d_L}
\end{eqnarray}
$m^2_{\tilde{d}_L}$ ($m^2_{\tilde{u}_L}$) are thereby the squared
masses of the down (up)-type squarks of the first two generations and
$(m_D^2)_{ij} ((m_U^2)_{ij})$ denotes the corresponding off-diagonal
elements of the squark mass matrices. The quantities $C_i$ and
$\tilde{N}$ are listed explicitly in the appendix.  

The detailed flavour structure of SUSY models is unknown and 
the R-parameters are thus complex, a priori unknown, quantities. It is
however possible to derive upper-limits using 
various experimental results\cite{BRS,CI,Bounds}. Since the analyses
of~\cite{BRS,CI} the experimental situation has not changed
significantly. Therefore our limits\footnote{ for the first two limits, 
thesign ``$\prec$" means that the upper limit of the real part (resp. the 
imaginary part) of the R-parameters  is the real part (resp. the imaginary part ) 
of the r.h.s.} , 
\begin{eqnarray*}
R^D_{s_L d_L} &\prec& (-105-56i)\frac{m_{\tilde{d_L}}}{500\; \mathrm{GeV}}
\\
R^U_{s_L d_L} &\prec& (-105-56i)\frac{m_{\tilde{u_L}}}{500\; \mathrm{GeV}}
\\
\mid R^U_{s_L t_R}\mid &<& \mathrm{Min}\left( 219\; 
(\frac{m_{\tilde{u_L}}}{500 
\; \mathrm{GeV}})^3,\ 43\right)
\\
\mid R^U_{t_R d_L}\mid &<& 41\,(\frac{m_{\tilde{u_L}}}{500\; \mathrm{GeV}})^2
\end{eqnarray*}
are very close to their limits. The main changes are due to the new
fits to the CKM matrix elements (central values). Considering the
errors in the CKM matrix elements and in the top quark mass, these
limits become slightly larger.  

Unfortunately the results for the branching ratio is very sensitive to
the SUSY parameters (masses and couplings) and only the future data of
the new generation experiments will give a significant improvement on
these quantities.  However, it is possible to estimate the order of
magnitude of the SUSY contributions.  The authors of Ref.~\cite{BRS} found for
$r_K$ and $\theta_K$ the typical ranges 
\begin{equation}
\label{rangeK} 
0.5<r_K<1.3,\ -25^o<\theta_K<25^o \; .
\end{equation}
by varying all SUSY parameters within the bounds allowed by
experimental constraints. This analysis is not restricted to some
specific model but it is valid for a general supersymmetric extension
of the standard model provided that new physics contributions to the
tree level decay $K^+\to \pi^0e^+\nu_e$ and contributions due to other
operators than $\bar{s_L}\gamma^{\mu}d_L\;
\bar{\nu^l_L}\gamma_{\mu}\nu^l_L$ can be neglected. One remark of
caution is in order here: these ranges are
not exclusive but they only indicate the most probable values.  Our
updated analysis agrees with this statement slightly enhancing the
probability for the values lying within the above range.

Varying $r_K$ and $\theta_K$ within these ranges makes at most a
change of $\sim$50\% for the branching ratio of $K^+\to\pi^+\nu\bar{\nu}$. 
We obtain
\[
BR^{Susy}(K^+ \to \pi^+ \nu \bar{\nu})=(8.18\ ^{+4.26}_{-5.23})
\times 10^{-11}~. 
\]
The central value thereby corresponds to the SM value, i.e. $r_K = 1$
and $\theta_K = 0$.  Allowing for wider, less probable ranges, we can
set an upper limit on the central value of approximately $1.55\times
10^{-10}$ (with $r_K=1.5$ and $\theta_K=-30^o$). Hence it is possible
to saturate the experimental central value with supersymmetric
contributions alone without any R-parity violating couplings.
 
In the same way as in the standard model case we can deduce branching
ratios for other rare kaon decays:
\begin{eqnarray} 
&&BR^{Susy}(K^0_L \to \pi^0 \nu \bar{\nu})=(2.90\ ^{+2.0}_{-2.18})
\times 10^{-11} \\
&&BR^{Susy}(K^0_S \to \pi^0 \nu \bar{\nu})=(0.57\ ^{+0.29}_{-0.39})
\times 10^{-12}
\end{eqnarray}
For the branching of $K^0_L \to \pi^0 e^+ e^-$ in new physics scenarios 
see~\cite{IsidoriKL2}.
The range given corresponds thereby to the range, Eq.~\ref{rangeK},
for the SUSY contributions and the central value corresponds to the SM
value. We would like to emphasize that these values can only be an
order of magnitude estimate since the precise value of the SUSY
corrections is not known. The range, Eq.~\ref{rangeK}, only gives the
most probable value, but is in no way exclusive. In addition, we did
not perform a really systematic analysis, including for instance the
errors on CKM parameters. However, these results show that the SUSY
corrections can be of the same order as the SM ones and that they
should thus be taken into account for an analysis of the constraints
imposed on R-parity violating couplings. The latter will be discussed
in more detail in the next section.

\section{The R-parity violating contribution}
\label{rparity}
\begin{figure}[ht]
\begin{center}
\epsfig{file=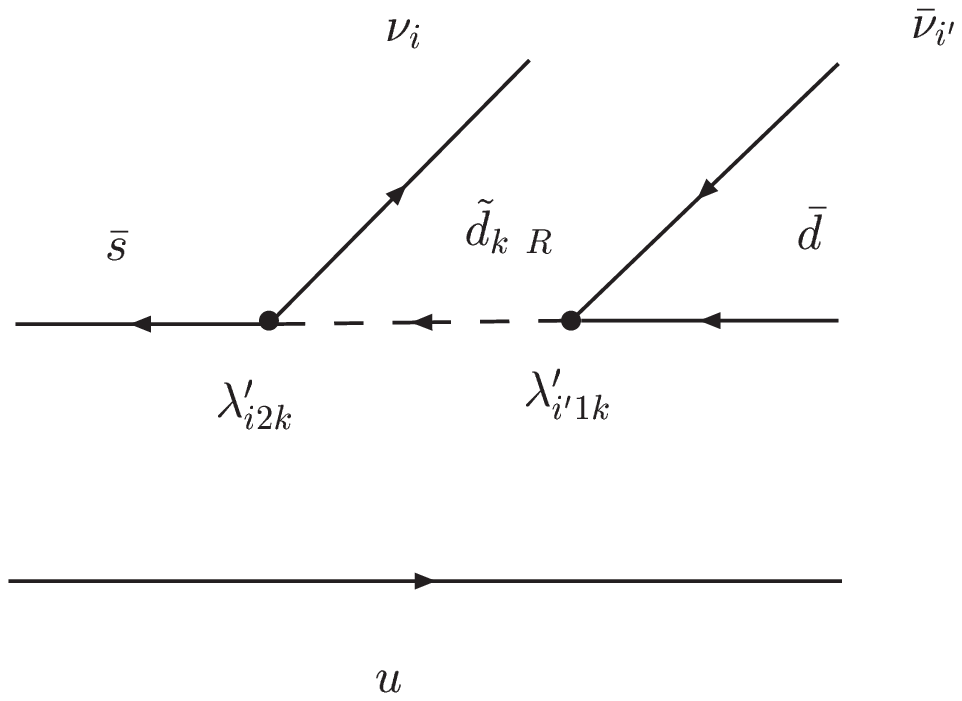,width=0.45\textwidth}
\epsfig{file=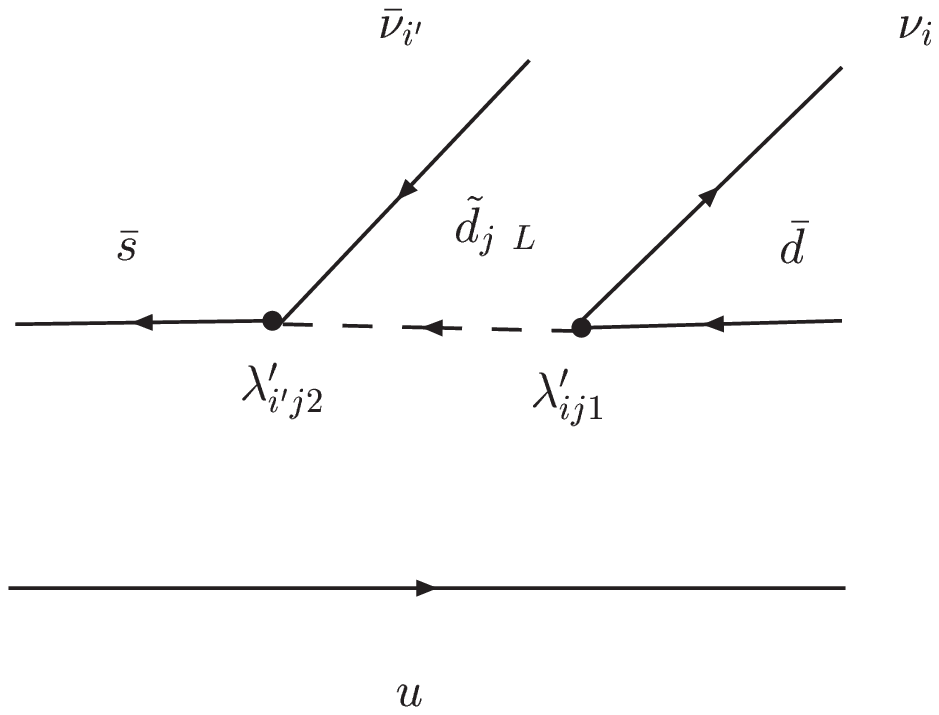,width=0.45\textwidth}
\caption{\label{fig1}\it R-parity violating diagrams contributing to 
the process $K^+ \to \pi^+ \nu{\bar{\nu}}$.}
\end{center}
\end{figure}

This section is devoted to a discussion of the contributions to the
process $K^+ \to \pi^+\nu\bar{\nu}$ arising
from R-parity violating couplings. Allowing for R-parity violation
in a supersymmetric extension of the standard model a superpotential
of the form
\begin{equation}
W = \lambda_{ijk} L_i L_j E_k + \lambda'_{ijk} L_i Q_j D^c_k +
\lambda^{''}_{ijk} U^c_i D^c_j D^c_k
\label{superpotential} 
\end{equation}
has to be considered. The first two
terms of this superpotential violate lepton number and the last one
violates baryon number. There are rather stringent limits on the
simultaneous presence of lepton and baryon number violating couplings
since they can mediate rapid proton decay. In this paper we will
concentrate on the $\lambda'_{ijk}$ couplings since they induce tree
level contributions via squark exchanges to the decay $K^+\to
\pi^+\nu\bar{\nu}$ (cf the diagrams shown in Fig.~\ref{fig1}).

We note that there is an ambiguity in defining the $\lambda'_{ijk}$
couplings because of an ambiguity in defining the lepton and Higgs
fields (for a more detailed discussion of that point see
Refs.~\cite{Davidson,PhysRep}). We will take the basis as defined by
the superpotential ({\it Super-CKM basis}), Eq.~\ref{superpotential}, 
and assign the CKM matrix to the up-type squarks.   

Including the R-parity violating processes, the branching ratio for 
the rare decay $K^+ \to \pi^+ \nu \bar{\nu}$ can be written in the 
following form:
\begin{equation} 
\label{KSM}
BR(K^+ \to \pi^+ \nu \bar{\nu})=\frac{\bar{\kappa}_+}{3\lambda^2}
\left( \sum_l \mid C_l^{\mathrm{model}} + \frac{\epsilon_{ll}}{4 k \, 
(200\; \mathrm{GeV})^2} \mid^2 + \sum_{b\not= l} \frac{\mid \epsilon_{bl} \mid^2}
{16 k^2 \,(200\; \mathrm{GeV})^4} \right) 
\end{equation}
Throughout our analysis we assume 
$C_l^{\mathrm{model}}$ to be either $C_l^{SM}= \lambda_cX_c^l +
\lambda_t\,X_t $ or $C_l^{SUSY} = \lambda_cX_c^l + \lambda_t\,
X_\mathit{new}$. 
The factor ``$k$'' is given by
\begin{equation} 
k=\frac{G_f \alpha (m_Z)}{\sqrt{2}\pi\sin^2(\theta_w)} = 8.88\times 10^{-8}\
\mathrm{GeV}^{-2}
\end{equation}
Errors are below 1\% and are therefore negligible compared to other errors.
The R-parity violating couplings are contained in the $\epsilon_{ij}$
which are defined as
\begin{equation} 
\epsilon_{ij}= \sum_n \left( \frac{\lambda_{i2n}^{'*}\lambda_{j1n}^{'}}
{m_{\tilde{d}n_R}^2} -\frac{\lambda_{in1}^{'*}\lambda_{jn2}^{'}}
{m_{\tilde{d}n_L}^2} \right) (200\;\mathrm{GeV})^2 \; .
\end{equation}
Note that we give our constraints for degenerate squarks masses of 200 GeV 
as the recent lower bounds on squarks have increased~\cite{PDG04}.
 $\lambda'_{ijk}$ are complex parameters such that the phase of
$\epsilon_{ij}$ is a priori not known.

In contrast to the standard model and the SUSY contributions, R-parity
violating couplings can induce processes with a neutrino and an
antineutrino of different flavour in the final state. This leads to the
last term in Eq.~\ref{KSM}. Obviously no interferences occur with the
standard model/SUSY contribution in that case. The R-parity violating
processes with the same neutrino flavour in the final state, however,
interfere with the SM/SUSY contribution as can be seen from the first
term in Eq.~\ref{KSM}.
The resulting contribution to the branching ratio is: 
\begin{equation} 
BR^{Int}=-2 \frac{\bar{\kappa}_+}{12\lambda^2 \; k (200\; \mathrm{GeV})^2 }
\sum_{l} Re(C_l^{model}\; \epsilon_{ll}) 
\end{equation}

%
\section{Full analysis and constraints on the $\lambda'$}
%
Within this section we will discuss different bounds on the
$\lambda'_{ijk}$ arising from the process $K^+\to \pi^+\nu\bar{\nu}$.
Of course it is not possible to establish bounds on each of the 27
couplings separately. Under some assumption it is possible to
constrain certain combinations of couplings.  We will perform our
analysis in three steps. First we will neglect all contributions
except the tree level R-parity violating ones in order to estimate
their order of magnitude in comparison with the SM and SUSY
contributions. The ``pure'' R-parity violating branching ratio is
given by
\begin{equation} 
BR^{R_p}= \frac{\bar{\kappa}_+}{48\lambda^2\;k^2 (200\; \mathrm{GeV})^4}
\sum_{i,j}
\mid \epsilon_{ij} \mid^2~. 
\end{equation}
Comparing this with the experimental value it is therefore straightforward to 
derive an upper bound for the sum
of all $\epsilon_{ij}$:
\begin{equation} 
\sum_{i,j} \mid \epsilon_{ij} \mid^2 < 5.6\times 10^{-10} \; .
\label{purerbound}
\end{equation}
With squarks at $100$ GeV, we have $\sum_{i,j} \mid \epsilon_{ij} \mid^2 <
0.35\times 10^{-10}$.
This bound is much lower than the upper bounds $2.3\times 10^{-9}$ and
$2.3\times 10^{-10}$ obtained respectively by 
Choudhoury and Roy \cite{CR} a few
years ago and recently by the authors of Ref.~\cite{PhysRep}, 
both using degenerate 
squarks masses of $100$ GeV. On the one hand this shows how much data 
on $K^+ \to \pi^+ \nu{\bar \nu}$ has improved recently, 
but on the other hand this implies
that the one-loop SM or SUSY contributions should be properly taken 
into account in order to obtain a realistic limit as they are of the
same order as the possible R-parity violating tree-level ones.

From our previous discussion, we have drawn the conclusion that SUSY
has to be included in the analysis of $K^+ \to \pi^+ \nu \bar{\nu}$.
In the next step we will therefore include the SUSY contribution. Note
that this contains the standard model contribution as a special case
($r_K = 1$ and $\theta_K = 0$). We will therefore not discuss the
latter separately. An analysis of the resulting bounds including only
the SM contribution has recently been performed for some special cases
in Ref.~\cite{DGHH04}.

Since we aim to obtain an upper-bound on the R-parity violating
couplings, we assume the SUSY contribution (which already includes SM)
to be minimal (corresponding to $r_K = 0.5$ and $\theta_K =
25^o$) in order to allow for the largest possible contribution
from R-parity violating terms. To simplify the discussion we will, for
the time being, neglect interferences (cf. section~\ref{rparity}). The
upper bound for the sum of the $\epsilon_{ij}$ can then slightly be
improved compared to the case without any SUSY contribution given in
Eq.~\ref{purerbound}. We obtain
\begin{equation} \sum_{i,j} \mid \epsilon_{ij} \mid^2 
< 4.45\times 10^{-10}~. 
\label{nointerfer}
\end{equation}
This limit can be translated to an upper bound on the product of two
couplings by naively setting all the couplings to zero except one
product. This procedure results in the following bounds: 
\begin{eqnarray} 
&&\mid\frac{\lambda_{i2n}^{'*}\lambda_{j1n}^{'}}
{m_{\tilde{d}n_R}^2} \mid\ < 2.11\times 10^{-5}
\label{produkt1} 
\\
&&\mid\frac{\lambda_{in1}^{'*}\lambda_{jn2}^{'}}
{m_{\tilde{d}n_L}^2} \mid\ < 2.11\times 10^{-5} 
\label{produkt2}
\end{eqnarray}

Of course, more realistic and precise constraints should take into
account interferences. This, however, makes the extraction of upper
bounds harder and no simple bounds in the sense of
Eqs.~\ref{nointerfer}, \ref{produkt1}, \ref{produkt2} can be given. In the following
part we will assume that only final states with the same neutrino
flavour occur. Thus only $\epsilon_{ij}$ with $i = j$ has to be taken
into account. The general equation verified by the $\epsilon_{ii}$ can
be written in the following way:
\begin{equation} 
\sum_{i=e,\mu,\tau}\left(Re(\epsilon_{ii}) + \frac{\alpha_i}{2}\right)^2\; +  
\sum_{i=e,\mu,\tau}\left(Im(\epsilon_{ii}) + \frac{\beta}{2}\right)^2=R^2\; ,
\label{cercleeq}
\end{equation}
Taking only one of the $\epsilon_{ii}$ nonzero, this equation
describes a circle in the complex plane, whose parameters are listed
in Table~\ref{table1} for the case of the standard model ($r_K = 1$ and
$\theta_K = 0$) and for the ``minimal'' SUSY contribution,
corresponding to $r_K = 0.5$ and $\theta_K = 25^o$. The radius R contains the
term $\displaystyle \frac{48\lambda^2\;k^2}{\bar{\kappa}_+ (200\; \mathrm{GeV})^{-4}}
(BR^{exp}-BR^{th})$ and 
the shifts $\alpha$ and $\beta$. The shifts depends on the CKM inputs and on the
loop-functions $X_t$ and $X_c^l$. The $c_i$'s are the corrections of $\alpha$ due to
the implicit dependence on the lepton masses $m_i$.

\begin{figure}[htb]
\begin{center}
  \mbox{\epsfxsize=0.6\textwidth
       \epsffile{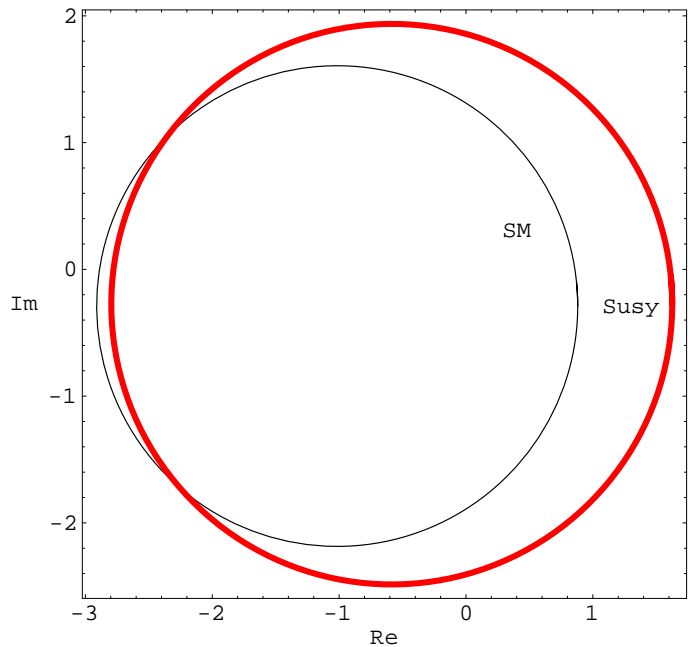}}
  \end{center}
\caption[]{\label{fig:plot1sigma} Allowed region for
$Re(\epsilon_{11})$ and $Im(\epsilon_{11})$ in units of
$10^{-5}$. In order to compare with the very recent constraints
on these R-parity violating couplings~\cite{DGHH04}, remember that we
take 200 GeV as reference value for the mass of the squarks.}
\end{figure}

The resulting constraints in the complex plane on $\epsilon_{11}$ are displayed in
Fig.~\ref{fig:plot1sigma}. Constraints on $\epsilon_{22}$ and
$\epsilon_{33}$ can be obtained in the same way and are of the same
order of magnitude.
To have an idea of the influence of the interferences, we may choose the point
of coordinates (Re($\epsilon_{11}$)=-2, Im($\epsilon_{11}$)=-2) on the SUSY
circle of Fig.2. It is approximately the point which gives the maximum value
for $\mid \epsilon_{11} \mid$. We have $\mid \epsilon_{11} \mid=2.8\times 10^{-5}$ 
and then we deduce, in the same
manner, limits on products of RPV couplings\footnote{ the case
i=$\tau$ gives slightly lower limits due to the correction factor
$c_{\tau}=0.81$, but the limits \ref{int-limits1}, \ref{int-limits2} can be
used for the 3 flavours.}:
\begin{eqnarray} 
&&\mid\frac{\lambda_{i2n}^{'*}\lambda_{i1n}^{'}}
{m_{\tilde{d}n_R}^2} \mid\ \lsim 2.8\times 10^{-5}
\label{int-limits1}
\\
&&\mid\frac{\lambda_{in1}^{'*}\lambda_{in2}^{'}}
{m_{\tilde{d}n_L}^2} \mid\ \lsim 2.8\times 10^{-5} 
\label{int-limits2}
\end{eqnarray}
These are 30$\%$ bigger than the ones in Eqs.~\ref{produkt1},~\ref{produkt2} and 
so this numerical example shows that interferences may have a significant 
influence. 

\TABLE[t]{
\begin{tabular}{l|c|c|c|c|c}
 & $\alpha_i$ &$c_e = c_\mu$& $c_\tau $& $\beta$ & $ R^2$ \\ \hline
SM & $c_i\,(2.03\times 10^{-5}) $
& 1& 0.89 &$ 5.79 \times 10^{-6}$ & 
$2.48\times 10^{-10}\, +
\sum_i\alpha_i^2/4 \,+\beta^2/4 $ \\
SUSY$_\mathrm{min}$ & $c_i\,(1.17\times 10^{-5})$ & 1 & 0.81 & 
$5.49 \times 10^{-6}$ & $4.47\times 10^{-10}\, +\sum_i\alpha_i^2/4 \,+\beta^2/4 $
\end{tabular}
\caption{Values of the constants appearing in Eq.~\ref{cercleeq}}
\label{table1}
}

%
\section{Conclusions}
%
We have investigated the decay $K^+ \to \pi^+ \nu {\bar \nu}$ and
related rare decays as a probe of physics beyond the standard
model. As a starting point we have obtained the standard model value
using updated experimental values for all the relevant parameters. We
found a slightly bigger branching than other recent estimates:
$B=(8.18 \pm 1.22)\times 10^{-11}$.  Furthermore we have analyzed the
supersymmetric contributions in the mass-insertion approximation and
corrected a minor misprint in the neutralino contribution that was
present in the existing literature. The main concern of this paper was
to obtain stringent limits on the R-parity violating
couplings. Assuming that the process $K^+\to \pi^+\nu{\bar\nu}$ is
governed entirely by RPV processes, the bounds on RPV couplings can be
lowered with respect to previous analysis
(cf. Refs.~\cite{CR,PhysRep}) due to recent data. Recent experimental
limits, however, indicate that SM and SUSY contributions, which can
give up to 50\% of the SM one, should be taken into account in order
to establish limits on the RPV couplings. We performed an analysis
including all these effects and we have shown that interferences can
have an effect of the order of 30 \% on the bounds on RPV couplings.

%
\section*{Acknowledgements}
%
We wish to thank G.~Isidori and the JHEP referee for their comments on the 
manuscript. We also thank M.~Bona, M.~Pierini and A.~Stocchi for providing the updated
values of the CKM fit, T.~Trippe and A.~Sher for mail exchange
concerning the recent data on the $K^+ \to \pi^0 e^+ \nu$ decay and finally 
S.~Davidson for a discussion on R-parity violation. Feynman diagrams are drawn 
using Jaxodraw~\cite{Binosi:2003yf}.

\appendix
%
\section{Explicit expressions of the supersymmetric contributions}
%
We only list the final formulae, more details and explanations (on
the basis of the fermions and sfermions fields, Feynman graphs)
can be found in Refs.~\cite{BRS,CI}. The misprints of \cite{BRS} have been 
corrected and functions are written in the notation of \cite{CI}.  We
based our calculations on~\cite{HK,IL}.

General notations :
\begin{itemize}
\item{$x_{ij}$ denote ratios of squared masses (for example : 
$\displaystyle x_{i\tilde{t}_R}=\frac{M_{\tilde{C}_i}^2}{M_{\tilde{t}_R}^2}$),}
\item{${\bf j}(x_1,..,x_n)$ and ${\bf k}(x_1,..,x_n)$ are the loop 
functions defined in \cite{BRS} 
(singularities of these functions in the case of equal arguments become derivatives).}
\end{itemize}

\subsection*{Charged Higgses contribution}

\begin{equation}X_{H^{\pm}}=X_{H^{\pm}}(x_H=\frac{M_W^2}{m_{H^\pm}^2})= \frac{m_t^2}{8 M_W^2 tan^2\beta} 
x_H\left( \frac{1}{(x_H-1)} -\frac{\ln(x_H)}{(x_H-1)^2} \right) \end{equation}

\subsection*{Chargino contributions}
\begin{itemize} 	
\item{$\displaystyle g_t=\frac{m_t}{\sqrt{2} M_W \sin\beta}$ is the top-quark coupling,}
\item{U and V are 2x2 unitary matrices that diagonalize the chargino mass matrix:

\begin{equation} M_{\tilde{C}}=U^T\,diag(M_{\tilde{C}_1},M_{\tilde{C}_2})\,V,\ M_{\tilde{C}_i}>0 \end{equation}

\begin{equation}M_{\tilde{C}}=\left(\begin{array}{cc}
M_2 & \sqrt{2} M_W \sin\beta \\
\sqrt{2} M_W \cos\beta & \mu
\end{array}\right) \end{equation}

They can be found in explicit form in \cite{diag1} \cite{diag2}.}
\end{itemize}

\subsubsection*{1-RR contribution}
\begin{equation}C^0_{RR}=\sum_{i;j=1,2} \left( (C^0)^{Pen}_{ij}+\ (C^0)^{Box}_{ij} \right)\end{equation}
\begin{eqnarray}
&&(C^0)^{Pen}_{ij}=\frac{g_t^2}{8}\,V_{2j}^{\dag}V_{i2}\left( {\bf k}(x_{i\tilde{t}_R},x_{j\tilde{t}R})\,V_{j1}V_{1i}^{\dag} 
-2\,{\bf
j}(x_{i\tilde{t}_R},x_{j\tilde{t}_R})\sqrt{x_{i\tilde{t}_R}\,x_{j\tilde{t}_R}}\,U_{i1}U_{1j}^{\dag}\right)\;
\\
&&(C^0)^{Box}_{ij}=\frac{g_t^2}{2}\,x_{W\tilde{t}_R}\,{\bf j}(x_{i\tilde{t}_R},x_{j\tilde{t}_R},x_{\tilde{e}_L\tilde{t}_R})\,V_{2j}^{\dag}V_{i2}U_{i1}U_{1j}^{\dag}
\end{eqnarray}

\subsubsection*{2-LL contribution}
\begin{equation}C_{LL}=\sum_{i;j=1,2} \left( (C_{LL})^{Pen}_{ij}+\ (C_{LL})^{Box}_{ij} \right)\end{equation}
				
\begin{eqnarray}  
(C_{LL})^{Pen}_{ij}&=&-\frac{1}{8}\,V_{1j}^{\dag}V_{i1}\left({\bf k}(x_{i\tilde{u}_L},x_{j\tilde{u}_L},1)\,V_{j2}V_{2i}^{\dag}  
- 2\,{\bf j}(x_{i\tilde{u}_L},x_{j\tilde{u}_L},1)\sqrt{x_{i\tilde{u}_L}\,x_{j\tilde{u}_L}}U_{i2}U_{2j}^{\dag}\right)
\nonumber\\ && \\
(C_{LL})^{Box}_{ij}&=&\frac{1}{2}\,x_{W\tilde{u}_L}\,{\bf j}(x_{i\tilde{u}_L},x_{j\tilde{u}_L},x_{\tilde{e}_L\tilde{u}_L},1)\,V_{1j}^{\dag}V_{i1}U_{i1}U_{1j}^{\dag}
\end{eqnarray}

\subsubsection*{3-LR and RL contribution}

\begin{equation}C_{LR}=\sum_{i;j=1,2} \left( (C_{LR})^{Pen}_{ij}+\ (C_{LR})^{Box}_{ij} \right)\end{equation}

\begin{equation}C_{RL}=\sum_{i;j=1,2} \left( (C_{RL})^{Pen}_{ij}+\ (C_{RL})^{Box}_{ij} \right)\end{equation}

\begin{eqnarray} 
(C_{LR})^{Pen}_{ij}&=&-\frac{m_t g_t}{8 m_{\tilde{u}_L}}\,V_{2j}^{\dag}V_{i1}\times 
\\ 
&&\left({\bf k}(x_{i\tilde{u}_L},x_{j\tilde{u}_L},x_{\tilde{t}_R\tilde{u}_L}) V_{j1}V_{1i}^{\dag}
-{\bf k}(x_{i\tilde{u}_L},x_{\tilde{t}_R\tilde{u}_L},1)\delta_{ij}\right.\nonumber \\ && \left. -2\,{\bf j}(x_{i\tilde{u}_L},x_{j\tilde{u}_L},x_{\tilde{t}_R\tilde{u}_L})U_{i1}U_{1j}^{\dag} 
\sqrt{x_{i\tilde{u}_L}\,x_{j\tilde{u}_L}}\right)
\nonumber
\\ 
(C_{LR})^{Box}_{ij}&=&-\frac{m_t g_t}{2
 m_{\tilde{u}_L}}\,x_{W\tilde{u}_L}\,{\bf j}(x_{i\tilde{u}_L},x_{j\tilde{u}_L},x_{\tilde{t}_R\tilde{u}_L},x_{\tilde{e}_L\tilde{u}_L})
 \,V_{2j}^{\dag}V_{i1}U_{i1}U_{1j}^{\dag}
\end{eqnarray}

\begin{eqnarray} 
(C_{RL})^{Pen}_{ij}&=&-\frac{m_t g_t}{8 m_{\tilde{u}_L}}\,V_{1j}^{\dag}V_{i2}\times
\\
&&\left( {\bf k}(x_{i\tilde{u}_L},x_{j\tilde{u}_L},x_{\tilde{t}_R\tilde{u}_L})\,V_{j2}V_{2i}^{\dag}
-{\bf k}(x_{i\tilde{u}_L},x_{\tilde{t}_R\tilde{u}_L},1)\delta_{ij}\right.\nonumber \\ &&\left. - 2\,{\bf j}(x_{i\tilde{u}_L},x_{j\tilde{u}_L},x_{\tilde{t}_R\tilde{u}_L})U_{i2}U_{2j}^{\dag} 
\sqrt{x_{i\tilde{u}_L}\,x_{j\tilde{u}_L}}\right)
\nonumber 
\\
(C_{RL})^{Box}_{ij}&=&-\frac{m_t g_t}{2
m_{\tilde{u}_L}}\,x_{W\tilde{u}_L}\,{\bf j}(x_{i\tilde{u}_L},x_{j\tilde{u}_L},x_{\tilde{t}_R\tilde{u}_L},x_{\tilde{e}_L\tilde{u}_L})V_{1j}^{\dag}V_{i2}U_{i2}U_{2j}^{\dag}
\end{eqnarray}

\subsection*{Neutralino contribution}

The neutralino contribution is potentially large since $R^D_{s_L d_L}$
can be large such that it should be taken into account.  Compared to
Ref.~\cite{BRS}, we corrected a minus sign in front of the function
{\bf k} in expression for the penguin diagram. An overall missing
factor (-1/2) has been corrected too.  The first two diagrams in
figure 7 in Ref. \cite{BRS} involving a squark-squark-$Z$ vertex are
cancelled by self-energy corrections. The two other contributions
give:
\begin{equation}
\tilde{N}=\sum_{n;m=1..4} \left( (\tilde{N})^{Pen}_{nm}+\ (\tilde{N})^{Box}_{nm} \right)
\end{equation}

\begin{eqnarray} 
(\tilde{N})_{nm}^{Pen}&=&
-\frac{1}{2}\,{\bf j}(x_{n\tilde{d}_L},x_{m\tilde{d}_L},1)\sqrt{x_{n\tilde{d}_L}\,x_{m\tilde{d}_L}}\,W^T_n(d_L)\lbrack W_{n4}W_{4m}^{\dag}-W_{n3}W_{3m}^{\dag}\rbrack W^*_m(d_L)
 \nonumber 
 \\
&&\ \ -\frac{1}{4}\,{\bf k}(x_{n\tilde{d}_L},x_{m\tilde{d}_L},1)\,W^T_n(d_L)\lbrack W^*_{n4}W_{4m}^T-W_{n3}^*W_{3m}^T\rbrack W^*_m(d_L)
\\ 
(\tilde{N})_{nm}^{Box}&=&x_{W\tilde{d}_L}\lbrace
{\bf k}(x_{n\tilde{d}_L},x_{m\tilde{d}_L},x_{\tilde{\nu}_L\tilde{d}_L},1)\,W^T_n(d_L)W^*_n(\nu_L) W^T_m(\nu_L)W^*_m(d_L)
\nonumber 
\\
&&\ \ +\,2\,{\bf j}(x_{n\tilde{d}_L},x_{m\tilde{d}_L},x_{\tilde{\nu}_L\tilde{d}_L},1)\sqrt{x_{n\tilde{d}_L}\,x_{m\tilde{d}_L}}\,W^T_n(d_L)W_n(\nu_L)
W^{\dag}_m(\nu_L)W^*_m(d_L)\rbrace \nonumber \\
\end{eqnarray}
W is the unitary matrix which diagonalizes the Neutralino mass 
matrix :
\begin{equation} M_{\tilde{N}}=W^T.diag(M_{\tilde{N}_1},M_{\tilde{N}_2},
M_{\tilde{N}_3},M_{\tilde{N}_4}).W\ \ \,, M_{\tilde{N}_n}>0
\end{equation}

\begin{equation} 
M_{\tilde{N}} = \left(\begin{array}{cccc}
M_1 & 0 & -M_Z \sin\Theta_W \cos\beta &  M_Z \sin\Theta_W \sin\beta \\
0 & M_2 &  M_Z \cos\Theta_W \cos\beta & -M_Z \cos\Theta_W \sin\beta \\
-M_Z \sin\Theta_W \cos\beta & M_Z \cos\Theta_W \cos\beta & 0 & -\mu \\
 M_Z \sin\Theta_W \sin\beta &-M_Z \cos\Theta_W \sin\beta & -\mu & 0
\end{array}\right)\nonumber
\end{equation}
By $W_n(x)$ we denote $T_3(x)\,W_{n2} + \tan\theta_W\,
\frac{Y(x)}{2}\,W_{n1}$, where $T_3(x)$ is the third component of the weak 
isospin and $Y(x)$ the hypercharge.


\end{document}